\UseRawInputEncoding
\documentclass[preprint,superscriptaddress,amsmath,amssymb,prd,aps,showpacs,floatfix,nofootinbib]{revtex4-1}
\usepackage{graphicx}
\usepackage{dcolumn}
\usepackage{bm}
\usepackage{mathrsfs}
\usepackage{hyperref}
\usepackage{amsmath}
\usepackage{amssymb}
\usepackage{bm}
\usepackage{color}
\usepackage{float}
\usepackage{dcolumn}
\usepackage{multirow}
\usepackage{changepage}
\usepackage{enumerate}
\hypersetup{pdftex,colorlinks=true,linkcolor=blue,citecolor=red,menucolor=black,urlcolor=blue,filecolor=blue}

\newcommand{\bl}[1]{\textcolor{blue}{#1}}
\newcommand{\re}[1]{\textcolor{red}{#1}}

\begin{document}

\title{The $D_s^+ \to \pi^+  K_s^0 K_s^0 $ reaction \\and the $I=1$ partner of the  $f_0(1710)$ state}
\date{\today}

\author{L.~R. Dai}
\email{dailianrong@zjhu.edu.cn}
\affiliation{School of Science, Huzhou University, Huzhou 313000, Zhejiang, China}
\affiliation{Departamento de F\'{\i}sica Te\'orica and IFIC, Centro Mixto Universidad de Valencia-CSIC Institutos de Investigaci\'on de Paterna, Aptdo.22085, 46071 Valencia, Spain}

\author{E.~Oset}
\email{oset@ific.uv.es}
\affiliation{Departamento de F\'{\i}sica Te\'orica and IFIC, Centro Mixto Universidad de
Valencia-CSIC Institutos de Investigaci\'on de Paterna, Aptdo.22085,
46071 Valencia, Spain}

\author{L. S. Geng}~\email{lisheng.geng@buaa.edu.cn}
\affiliation{School of Physics, Beihang University, Beijing 102206, China}

\begin{abstract}
We have identified the decay modes of the  $D_s^+ \to  \pi^+ K^+ K^- , \pi^+ K_s^0 K_s^0$  reactions producing two
vector mesons and one pseudoscalar. The posterior vector-vector interaction generates two resonances that we associate
to the $f_0(1710)$ and the  $a_0(1710)$ recently claimed. We find two acceptable scenarios that give results for the ratio
of the branching ratios of these two reactions in  agreement with experiment. With these two scenarios we make predictions
for the branching ratios of the $D_s^+ \to \pi^0 K^+ K_s^0$ reaction, finding values within the range of $(2.0 \pm 0.7)\times 10^{-3}$.
Comparison of these predictions with coming experimental results on that latter reaction will be very most useful to deepen
our understanding on the nature of these two resonances.
\end{abstract}

\maketitle

\section{Introduction}
\label{sec:intro}
The $f_0(1710)$ is a well established meson in the PDG \cite{pdg}. In the relativized  quark model of  Godfrey and Isgur  \cite{isgur} it appears as an $I=0,J^{PC}=0^{++}$ state at $1780$ MeV with the $2$\,$^3P_1$ configuration.
In the same work a state with the same mass and configuration appears for $I=1$. Similar results are also reported  in \cite{entem}. The $f_0(1710)$ is also obtained in \cite{vijande} with the same configuration, but
no mention is made of the possible $I=1$ partner. These models consider the excitation of $u,d$ quarks. However, the fact that the $f_0(1710)$  decays mostly in $K\bar{K}, \eta\eta$ with only about $4\%$ branching ratio to
$\pi\pi$ decay \cite{pdg} indicates that this state should have large components of $s\bar{s}$ quarks.

 A different picture for the $f_0(1710)$ comes from the work of  \cite{gengvec}, where the interaction  of vector mesons studied  in \cite{raquelvec} for the $\rho\rho$ case is extended  to the SU(3) space. The
 $f_0(1710)$ is found around $1726$ MeV and couples mostly to $K^* \bar{K}^*$, but also to $\omega\phi$, $\phi\phi$, $\omega\omega$, $\rho\rho$ in that order.
 The vector-vector interaction is taken from the local hidden gauge approach \cite{hidden1,hidden2,hidden4,hideko}, which stems from a contact term plus vector exchange.
 Considering box diagrams  with intermediate two pseudoscalar mesons, decay rates to $K\bar{K}, \eta\eta, \pi\pi$  were evaluated in \cite{gengvec} and found consistent with experimental data. Interestingly, in
 \cite{gengvec} a parter state of the $f_0(1710)$ with $I=1, J^{PC}=0^{++}$ $a_0$ state is also found at $1780$ MeV with $\Gamma \sim 130$ MeV. This
 state couples mostly to $K^* \bar{K}^*$ but also to $\rho\omega$ and $\rho\phi$.  The picture of \cite{gengvec} for the $f_0(1710)$  has been tested in different processes. In \cite{genghideko}
 the $\gamma\gamma$ decay rate is evaluated and found consistent with the PDG information  \cite{pdg}. In  \cite{albeto}  it is also suggested that the peak observed  at the  $\phi\omega$
 threshold in the $\phi\omega$  mass distribution of the $J/\psi \to \gamma \phi\omega $ decay \cite{expfiom} is due to the  $f_0(1710)$  resonance. Predictions for other decay modes, and rates for $f_0(1710)$
  production decays of other particles are done in \cite{gengguo,yamagata,jujun,daijujun,daigeng,ikenoliang}.

  The success of the predictions for other  vector-vector molecules obtained in \cite{gengvec} discussed in the former references gives us confidence in that  model and concretely  about the existence  of the $I=1$
  partner of the $f_0(1710)$  state, which we will  call the  $a_0(1710)$ by analogy to the $f_0(1710)$. Yet, this state  is not reported in the PDG \cite{pdg}. The situation has changed suddenly with the
  appearance of two works showing evidence  for this state. One of these works is the clear observation of a peak  around $1710$ MeV  in the $\pi^+ \eta$ mass distribution in the
  $\eta_c \to \eta \pi^+  \pi^-$ decay \cite{lees}.  The other work  is the study of  the $D_s^+ \to \pi^+  K_s^0 K_s^0 $ decay  \cite{besnew} showing a peak around $1710$ MeV   in the  $K_s^0 K_s^0$ mass distribution with
  an abnormally large strength  compared to the one  of a similar peak seen in the   $D_s^+ \to  \pi^+  K^+ K^-$  decay \cite{besold}. This cannot be explained from a $K\bar{K}$ state in $I=0$, implying that
  there must also  be an $I=1$ state with a similar mass.  The $I=0$ and $I=1$ states have relative opposite sign in the $K^+ K^- $ or $K^0 \bar{K}^0$ components and the contribution of the two states
  around $1710$ MeV  will give differences in the $K^+ K^-$  or  $K^0 \bar{K}^0$ production rates.

    Our aim in the present work is to show that from the perspective of Ref. \cite{gengvec} for the $f_0(1710)$ and $a_0(1710)$ it is natural to reproduce  the experimental data  on $K\bar{K}$ production in
   $D_s^+ \to \pi^+ K \bar{K} $ decay. At the same time we  can make  predictions for the rate of the $I=1$  $a_0(1710)$ production in the $K^+ K_s^0 $ invariant mass distribution of the
   $D_s^+ \to \pi^0 K^+ K_s^0  $ reaction, which, as mentioned in \cite{besnew} is in the process of being analyzed at BESIII.

\section{FORMALISM}

\begin{figure}[h]
\centering
\includegraphics[scale=0.95]{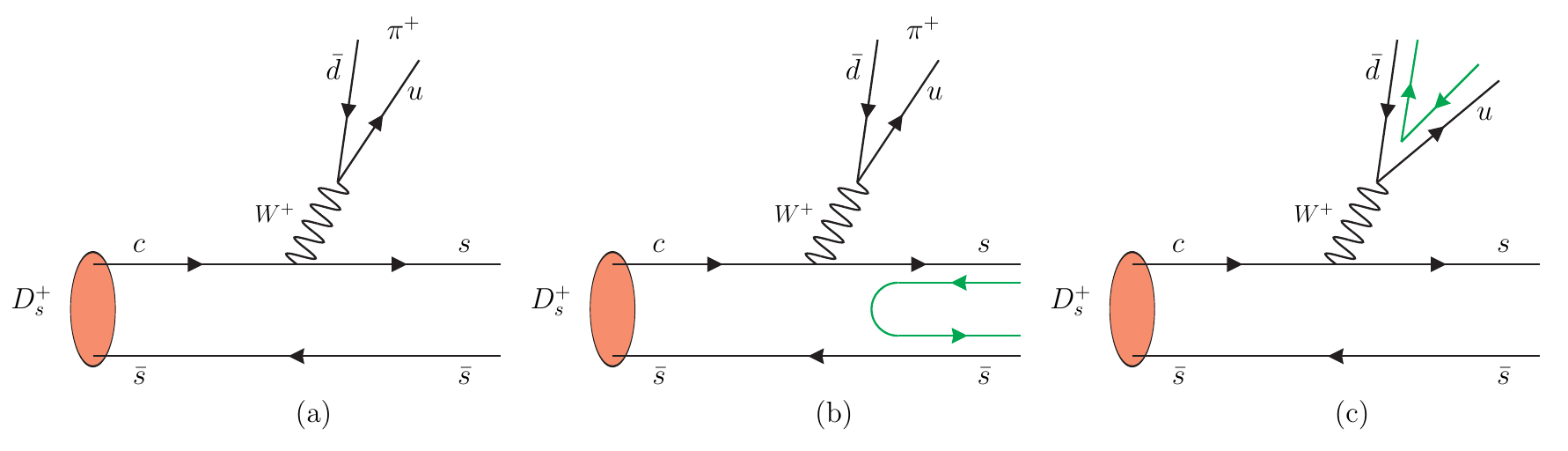}
\caption{Cabibbo-favored decay mode of  $D_s^+$ at the quark level with external emission (a); Hadronization  of the $s\bar{s}$ component (b) ; Hadronization  of the $\bar{d}u$ component (c).}
\label{fig:cabi1}
\end{figure}

 We  look at the mechanism for $D_s^+ \to  \pi^+ K \bar{K} $ production at the quark level  starting from the dominant external emission process and then considering the internal emission, both in the Cabibbo-favored
 mode \cite{chau}. The external emission with $\pi^+$  production is shown in Fig.~\ref{fig:cabi1}.

Since  we wish to have three mesons in the final state we must hadronize a pair of quarks introducing an extra $\bar{q}q$  with the vacuum quantum numbers $(\bar{q}q=\bar{u}u+\bar{d}d+\bar{s}s)$.
Also, the hadronization  must produce a pair of vector mesons, such that their interaction can produce the  $f_0(1710)$  and  $a_0(1710)$ resonances.
Then, hadronizing $s\bar{s}$ we will have
\begin{eqnarray}
  s\bar s \to \sum_i s\, \bar q_i q_i \, \bar s = V_{3i} V_{i3} = (V^2)_{33},
\end{eqnarray}
where $V$ is the $q_i \bar q_j$ matrix written in terms of the vector meson
\begin{eqnarray}\label{eq:V}
V = \left(
  \begin{array}{ccc}
 \frac{\rho^0 }{\sqrt{2}}+ \frac{\omega}{\sqrt{2}}  & \rho^+ & K^{*+} \\[2mm]
   \rho^- & -\frac{\rho^0 }{\sqrt{2}}+ \frac{\omega}{\sqrt{2}}  &K^{*0} \\[2mm]
  K^{*-} & \bar{K}^{*0} & \phi\\
     \end{array}
    \right)
\end{eqnarray}

Then we have the hadronic state
\begin{eqnarray}
H_1=(V^2)_{33}\pi^+=( K^{*-} K^{*+} +  \bar{K}^{*0} K^{*0} +\phi\phi) \, \pi^+ \,.
\end{eqnarray}

With the implicit isospin phase convention of this matrix, with the $(K^{*+},K^{*0})$, $(\bar{K}^{*0}, -K^{*-})$ isospin doublets, the former
combination represents $K^* \bar{K}^*$ in isospin $I=0$, as it should be, plus $\phi\phi$, also $I=0$, since originally we had the $I=0$ $s\bar s$ state.

  The two isospin states of $K^* \bar{K}^*$ are given by
\begin{eqnarray}\label{eq:wfnKs}
|K^* \bar{K}^*, I=0\rangle &=& -\frac{1}{\sqrt{2}} \big(K^{*+} K^{*-} \, + \, K^{*0}\bar{K}^{*0} \big) \nonumber  \\
|K^* \bar{K}^*, I=1,I_3=0\rangle &=& -\frac{1}{\sqrt{2}} \big(K^{*+} K^{*-} - K^{*0}\bar{K}^{*0} \big)
\end{eqnarray}

 We could also think of hadronizing  the $u\bar{d}$ pair with $VV$ but then $s\bar{s}$ should be a pseudoscalar, in this case a combination
 of  $\eta$ and $\eta'$, and we do not get the $\pi^+ K \bar{K}$ mode.
 Note that if we wish to have  $\pi^+ K \bar{K}$, we  could also have the $s\bar{s}$ as the $\phi$ meson and then $\phi \to K \bar{K}$, but
 the invariant mass of $K \bar{K}$ will peak at the $\phi$ mass and we are only concerned about the vicinity of $1710$ MeV, where the
  $f_0(1710)$  and  $a_0(1710)$ resonances appear. This decay mode and the $K^+ K^- \pi^+$  spectrum at low $K^+ K^-$ invariant mass has been
  studied in detail in \cite{xiao}. We shall concentrate here only in the region of  $f_0(1710)$  and  $a_0(1710)$ production.

 We have then another possibility which is to hadronize the $u\bar{d}$ component with $VP$ or $PV$ ($P$ for pseudoscalar meson).
 Similarly to Eq.~\eqref{eq:V} we have the $q_i \bar q_j$  matrix for pseudoscalars
 \begin{eqnarray}\label{eq:P}
P = \left(
  \begin{array}{ccc}
 \frac{\pi^0 }{\sqrt{2}}+ \frac{\eta}{\sqrt{3}}  & \pi^+ & K^+ \\[2mm]
   \pi^- & -\frac{\pi^0 }{\sqrt{2}}+ \frac{\eta}{\sqrt{3}}  & K^0 \\[2mm]
  K^- & \bar{K}^0 & -\frac{\eta}{\sqrt{3}}\\
     \end{array}
    \right)
\end{eqnarray}
where we have used the $\eta$ and $\eta'$ mixing of Ref.~\cite{bramon} and neglected the $\eta'$  which does not play any role here.

In this case we obtain the contribution
\begin{eqnarray}
u\bar d \to \sum_i u\, \bar q_i q_i \, \bar d = M_{1i} M'_{i2} = (MM')_{12}
\end{eqnarray}
but now $M,M'$ can be vector or pseudoscalar. Hence, we obtain the combinations
\begin{eqnarray} \label{eq:VP12}
(VP)_{12}=\big(\frac{\rho^0 }{\sqrt{2}}+ \frac{\omega}{\sqrt{2}} \big)\,\pi^+  \,+\, \rho^+ \big( \frac{-\pi^0 }{\sqrt{2}}+ \frac{\eta}{\sqrt{3}}\big)+ K^{*+}\bar{K}^0
\end{eqnarray}
\begin{eqnarray}\label{eq:PV12}
(PV)_{12}=\big( \frac{\pi^0 }{\sqrt{2}}+ \frac{\eta}{\sqrt{3}}\big) \rho^+  \,+\, \pi^+ \big(\frac{-\rho^0 }{\sqrt{2}}+ \frac{\omega}{\sqrt{2}} \big) + K^+ \bar{K}^{*0}
\end{eqnarray}
and the $s\bar{s}$ pair will provide the $\phi$ meson.

We aim at getting $\pi^+ f_0(1710)$ and $\pi^+ a_0(1710)$ which have $G$-parity negative and positive respectively. Neither $VP$ or $PV$ of Eqs.~\eqref{eq:VP12},~\eqref{eq:PV12}
have good $G$-parity  but the combinations $VP \pm PV$ have. Thus, we construct
\begin{eqnarray}\label{eq:H2}
H_2=\phi [(VP)_{12}+(PV)_{12}]= \big[2 \frac{\omega}{\sqrt{2}}\pi^+ \,+\,\frac{2}{\sqrt{3}}\rho^+ \eta \,+\, K^{*+}\bar{K}^0 \,+\, K^+ \bar{K}^{*0} \big]\phi
\end{eqnarray}
\begin{eqnarray}\label{eq:H3}
H_3=\phi [(VP)_{12}-(PV)_{12}]= \big[2 \frac{\rho^0}{\sqrt{2}}\pi^+ \,-\,\frac{2}{\sqrt{2}}\rho^+ \pi^0 \,+\, K^{*+}\bar{K}^0 \,-\, K^+ \bar{K}^{*0} \big]\phi
\end{eqnarray}
The combination of Eq.~\eqref{eq:H2} has $G$-parity  positive while the case of Eq.~\eqref{eq:H3} has $G$-parity negative\footnote{To facilitate testing the  $G$-parity
of the $K$, $K^*$ states we note that we have $G K^+=\bar{K}^0$, $G K^0=-K^-$, $G \bar{K}^0=-K^+$, $G K^-=K^0$, and the same with a global minus sign for $K^*$, since
we have the convention $C \rho^0=-\rho^0$, for the charge conjugation of vector mesons.}. With the latter combination we are able to reach the $\pi^+ f_0(1710)$  state,
while from  Eq.~\eqref{eq:H2} we can produce $\pi^+ a_0(1710)$.

\begin{figure}[h]
\centering
\includegraphics[scale=1.]{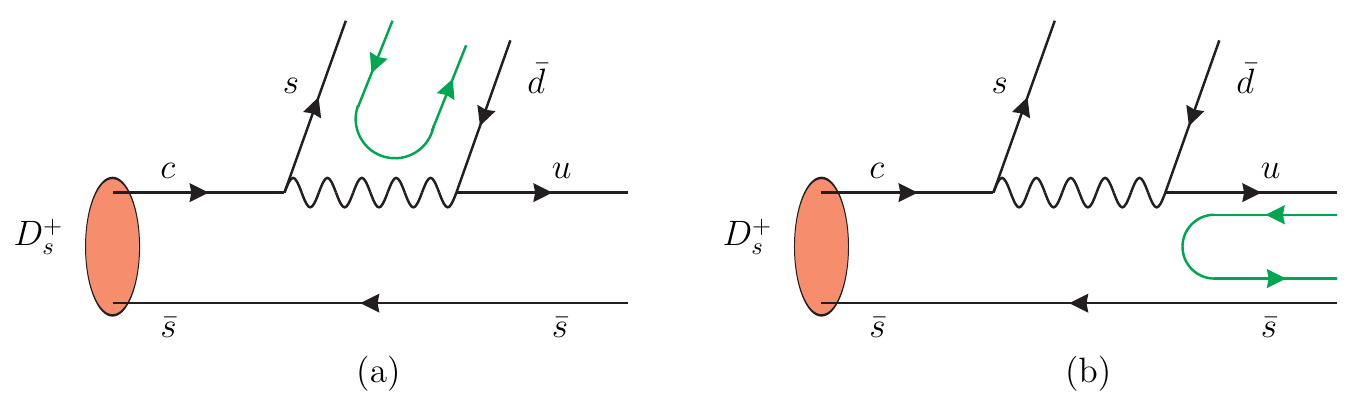}
\caption{Internal emission. (a) with  hadronization of the  $s\bar{d}$  pair;~~(b) with  hadronization of the  $u\bar{s}$  pair.}
\label{fig:cabi2}
\end{figure}

So far we have relied upon external emission. Suppressed by a color factor $\frac{1}{N_c}$ we have internal emission, which is depicted in Fig.~\ref{fig:cabi2}. We could
hadronize the $s\bar{d}$ or $u\bar{s}$ components with $VV$ but then we neither get a pion nor the $VV$ combination to give the nonstrange  $f_0(1710)$ or  $a_0(1710)$.
We must  hadronize with $VP$ and $PV$ combinations and we get
\begin{eqnarray}
(VP)_{32}&=& K^{*-}\pi^+ \,+\, \bar{K}^{*0} \big( \frac{-\pi^0 }{\sqrt{2}}+ \frac{\eta}{\sqrt{3}}\big)  + \phi \bar{K}^0  \nonumber  \\
(PV)_{32}&=& K^-\rho^+  \,+\, \bar{K}^0 \big(\frac{-\rho^0 }{\sqrt{2}}+ \frac{\omega}{\sqrt{2}} \big)-\frac{\eta}{\sqrt{3}} \bar{K}^{*0}    \nonumber  \\
(VP)_{13}&=& \big(\frac{\rho^0 }{\sqrt{2}}+ \frac{\omega}{\sqrt{2}} \big) K^+ \,+\, \rho^+ K^0- K^{*+}\frac{\eta}{\sqrt{3}} \nonumber  \\
(PV)_{13}&=&  \big( \frac{\pi^0 }{\sqrt{2}}+ \frac{\eta}{\sqrt{3}}\big)  K^{*+}  \,+\,   \pi^+ K^{*0} \,+\,  K^+\phi
\end{eqnarray}

 We must form the good $G$-parity combination from the former terms and we find
\begin{eqnarray}
H_4&=& K^{*+} (VP)_{32}+ \bar{K}^{*0} (PV)_{13}  \nonumber  \\
&=& \pi^+ (K^{*+} K^{*-} + \bar{K}^{*0}K^{*0}) \,+\,\frac{2}{\sqrt{3}} \eta  K^{*+} \bar{K}^{*0} + \phi (K^{*+}\bar{K}^0+ \bar{K}^{*0}K^+)
\end{eqnarray}
\begin{eqnarray}
H_5&=& K^{*+} (VP)_{32}- \bar{K}^{*0} (PV)_{13}  \nonumber  \\
&=& \pi^+ (K^{*+} K^{*-} - \bar{K}^{*0}K^{*0}) \,-\,\sqrt{2} \pi^0  K^{*+} \bar{K}^{*0} + \phi (K^{*+}\bar{K}^0 - \bar{K}^{*0}K^+)
\end{eqnarray}
Note that $(PV)_{32}$ and $(VP)_{31}$ combinations have no pions and do not lead to our desired  final state. We see again that $H_4$ has $G$-parity
negative and can lead to $\pi^+ f_0(1710)$, while $H_5$ has $G$-parity positive and can lead to $\pi^+ a_0(1710)$. The different mechanisms  have
different weights and we shall give weights:
$$
H_1:A \, \qquad H_2:A \alpha \, \qquad  H_3:A \beta \, \qquad  H_4:A \gamma \, \qquad H_5:A \delta \,.
$$
We will evaluate ratios of  $\pi^+ K_s^0 K_s^0 $  and $\pi^+ K^+ K^-$  production and the global factor $A$ disappears. Then we have $4$ parameters to adjust
an experimental ratio, which seems an excessive freedom, but we are very limited since $|\alpha|\sim 1$,  $|\beta|\sim 1$, $|\gamma| \sim \frac{1}{3}$, $|\delta|\sim \frac{1}{3}$
and then the freedom is drastically reduced.

The hadronic states $H_i$ $(i=1,2,3,4,5)$ do not have $K\bar{K}$ in the final state. We must produce the $f_0(1710)$ and $a_0(1710)$ and then let them decay into $K\bar{K}$ .
The mechanisms for $K\bar{K}$ decay are explained in \cite{gengvec},  but since we only care about ratios, all that is needed are the Clebsch-Gordan coefficients, and
considering the wave functions ($|K\bar{K},I=0\rangle =-\frac{1}{\sqrt{2}}(K^+ K^- +K^0 \bar{K}^0), |K\bar{K},I=1,I_3=0\rangle =-\frac{1}{\sqrt{2}}(K^+ K^- -K^0 \bar{K}^0),
~|K\bar{K},I=1,I_3=1\rangle =K^+\bar{K}^0$), we will have the weights
\begin{eqnarray}\label{eq:wf}
&f_0(1710) \to  &
      \left\{
    \begin{array}{ll}
    K^+ K^- =-\frac{1}{\sqrt{2}} g_{K\bar{K}}   \\[2mm]
    K^0 \bar{K}^0 =-\frac{1}{\sqrt{2}} g_{K\bar{K}}  \\[2mm]
    \end{array}
   \right.   \nonumber  \\
&a_0(1710) \to  &
\left\{
    \begin{array}{ll}
    K^+ K^- =-\frac{1}{\sqrt{2}} g_{K\bar{K}} \,, &I_3=0 \\[2mm]
    K^0 \bar{K}^0 =\frac{1}{\sqrt{2}} g_{K\bar{K}} \,, &I_3=0\\[2mm]
    \end{array}
   \right.  \nonumber  \\
&a_0(1710) \to & K^+ \bar{K}^0 =g_{K\bar{K}}\,,~~I_3=1
\end{eqnarray}
 and we also need the couplings of  the resonances to the different vector-vector channels, which we take from \cite{gengvec} and
 show in Table \ref{table:res1}.

 \begin{table*}[htpb]
 \renewcommand{\arraystretch}{1.0}
  \setlength{\tabcolsep}{0.1cm}
 \centering
 \caption{Couplings of  $f_0(1710)$  and  $a_0(1710)$  to $VV$ channels.  All quantities are in units of MeV.\label{table:res1}}
     \begin{tabular}{c|ccccc}
     \hline\hline
    & ~~ $K^*\bar{K}^*$ ~~&~~ $\rho\rho$~~ & ~~$\omega\omega$~~  &~~ $\omega\phi$~~ &~~ $\phi\phi$ \\
 $g[f_0(1710)]$~~ &  ~$(7124,i96)$ ~& ~$(-1030,i1086)$ ~& ~$(-1763,i108)$~ &~ $(3010,-i210)$~ &~ $(-2493,-i204)$\\
    \hline
     & ~~ $K^*\bar{K}^*$ ~~&~~ $\rho\rho$ ~~&~~ $\rho\omega$ ~~ &~~ $\rho\phi$\\
$g [a_0(1710)]$~~ &~~  $(7525,-i1529)$ & $0$ & $(-4042,i1391)$ & ~~$(4998,-i1872)$  \\
  \hline\hline
    \end{tabular}
 \end{table*}

\begin{figure}[h!]
\centering
\includegraphics[scale=0.8]{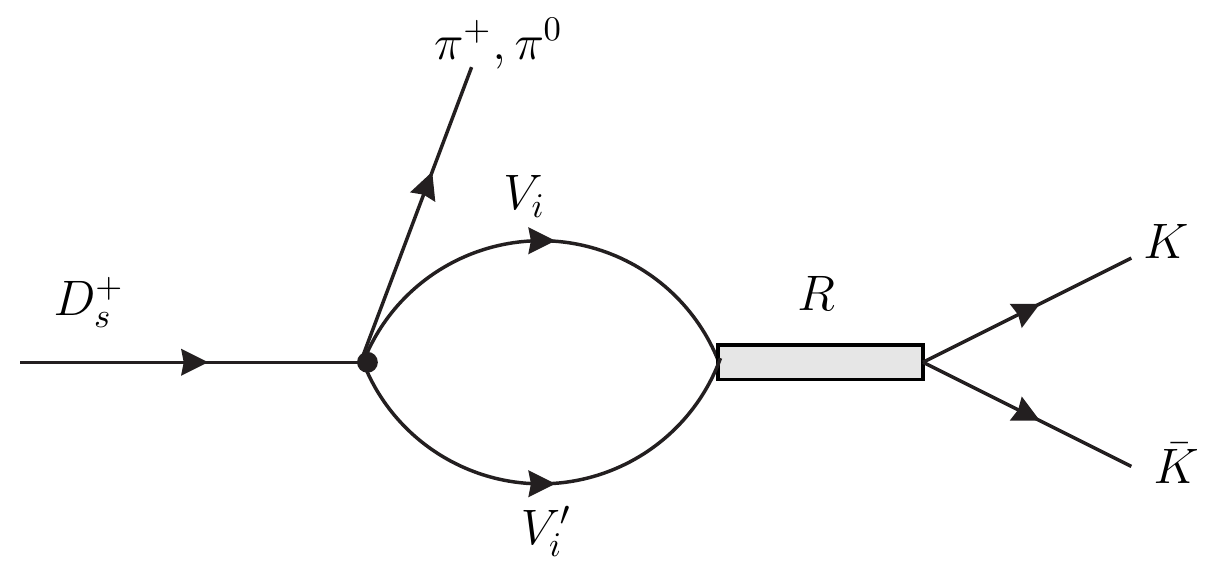}
\caption{Mechanisms for $D_s^+ \to  \pi^+ K^+ K^- (K^0 \bar{K}^0)$ and  $D_s^+ \to \pi^0 K^+ \bar{K}^0 $. For $\pi^+ f_0(1710)$ production  $V_i V'_i\equiv K^*\bar{K}^*,\omega\phi,\phi\phi$; for  $\pi^+ a_0(1710) ~(I_3=0)$ production $V_i V'_i\equiv K^*\bar{K}^*,\rho^0 \phi$;~
 for  $\pi^+ a_0(1710) ~(I_3=1)$ production $V_i V'_i\equiv K^*\bar{K}^*,\rho^+ \phi$. }
\label{fig:R}
\end{figure}

 We can see how the different $H_i$ terms contribute to $\pi^+ f_0(1710)$, $\pi^+ a_0(1710) (I_3=0)$ and $\pi^+ a_0(1710) (I_3=1)$.

  \begin{eqnarray}
\left\{
    \begin{array}{ll}
    H_1:& \pi^+ f_0(1710) ~{\rm with} ~\pi^+ K^*\bar{K}^* {~\rm and~} \phi\phi {~\rm terms}.  \\[2mm]
    H_2:& \pi^+ f_0(1710) ~{\rm with} ~ \omega\phi \pi^+ {~\rm term}. \\[2mm]
    H_3:& \pi^+ a_0(1710)~(I_3=0)~{\rm with} ~ \pi^+ \rho^0 \phi  {~\rm term}; \\[2mm]
    & \pi^+ a_0(1710)~(I_3=1)~{\rm with} ~ \pi^0 \rho^+ \phi  {~\rm term}. \\[2mm]
    H_4:& \pi^+ f_0(1710) ~{\rm with~} \pi^+ K^* \bar{K}^* {~\rm term}. \\[2mm]
    H_5:& \pi^+ a_0(1710)~(I_3=0)~{\rm with} ~ \pi^+ K^* \bar{K}^*  {~\rm term}; \\[2mm]
    & \pi^+ a_0(1710)~(I_3=1)~{\rm with} ~ \pi^0 K^* \bar{K}^*  {~\rm term}.
    \end{array}
   \right.
\end{eqnarray}
The mechanism for  $f_0(1710)$  and  $a_0(1710)$  production and $K\bar{K}$ final state  are  depicted in Fig.~\ref{fig:R}.

 All this said, and with  the weights of the different  mechanisms we can write
\begin{eqnarray}\label{eq:f0}
 {\tilde{t}_{f_0}} &=& A \lbrace -\sqrt{2}\,(1+ \re{\gamma})\, G_{K^* \bar{K}^*}(M_{\rm inv})\, g_{f_0,K^* \bar{K}^*}
 + 2 \,\times \frac{1}{2} G_{\phi\phi}(M_{\rm inv})\sqrt{2}\, g_{f_0,\phi\phi}  \nonumber   \\
&+&  \sqrt{2}\,\bl{\alpha} \,G_{\omega \phi}(M_{\rm inv})\, g_{f_0,\omega\phi}\rbrace
\end{eqnarray}
\begin{eqnarray}\label{eq:a01}
{\tilde{t}_{a_0}}(I_3=0) &=& A \lbrace \sqrt{2}\,\bl{\beta}\, G_{\rho\phi}(M_{\rm inv})\, g_{a_0,\rho\phi} -\sqrt{2}\,\re{\delta} \,G_{K^* \bar{K}^*}(M_{\rm inv})\, g_{a_0,K^* \bar{K}^*}\rbrace
\end{eqnarray}
\begin{eqnarray}\label{eq:a02}
{\tilde{t}_{a_0}}(I_3=1) &=& A \lbrace \sqrt{2}\,\bl{\beta}\, G_{\rho\phi}(M_{\rm inv})\, g_{a_0,\rho\phi} -\sqrt{2}\,\re{\delta} \,G_{K^* \bar{K}^*}(M_{\rm inv})\, g_{a_0,K^* \bar{K}^*}\rbrace
\end{eqnarray}
where we have taken into account the $K^* \bar{K}^*$ wave functions of Eqs.~\eqref{eq:wfnKs}, and, considering the isospin multiplet $(-\rho^+,\rho^0,\rho^-)$, the $\rho\phi$ wave functions
\begin{eqnarray}
|\rho\phi; I=1,I_3=0\rangle&=&\rho^0 \phi \nonumber   \\
|\rho\phi; I=1,I_3=1\rangle&=& -\rho^+ \phi \,.
\end{eqnarray}

The $G$ functions  in Eqs.~\eqref{eq:f0}, \eqref{eq:a01},~\eqref{eq:a02} are the loop functions for pairs of vector mesons, which are calculated using a cutoff method with $q_{\rm max} =960$ MeV,
giving similar results as those found in \cite{gengvec}, where dimensional regularization was used.

Since in all mechanisms we started from $s\bar{d}$, $u\bar{s}$, this state is $I=1,I_3=1$. Considering the phase of $\pi^+$,  the components $\pi^+ a_0$ $(I_3=0)$ and $\pi^0 a_0$ $(I_3=1)$
have the same weights in $I=1,I_3=1$, which means that Eqs.~\eqref{eq:a01},~\eqref{eq:a02} should be the same, which is indeed the case.

Considering the weights of  Eqs.~\eqref{eq:wf} of the resonances to $K\bar{K}$, we can then write:

\begin{eqnarray}\label{eq:ti}
t_{K^+ K^-} &=& -{\tilde{t}_{f_0}} \frac{1}{M^2_{\rm inv}-M^2_{f_0}+i M_{f_0} \Gamma_{f_0}} \, \frac{1}{\sqrt{2}}\,g_{K\bar{K}}
-{\tilde{t}_{a_0}} \frac{1}{M^2_{\rm inv}-M^2_{a_0}+i M_{a_0} \Gamma_{a_0}} \, \frac{1}{\sqrt{2}}\,g_{K\bar{K}} \nonumber   \\
t_{K^0 \bar{K}^0} &=& -{\tilde{t}_{f_0}} \frac{1}{M^2_{\rm inv}-M^2_{f_0}+i M_{f_0} \Gamma_{f_0}} \, \frac{1}{\sqrt{2}}\,g_{K\bar{K}}
+ {\tilde{t}_{a_0}} \frac{1}{M^2_{\rm inv}-M^2_{a_0}+i M_{a_0} \Gamma_{a_0}} \, \frac{1}{\sqrt{2}}\,g_{K\bar{K}} \nonumber   \\
t_{K^+ \bar{K}^0} &=&{\tilde{t}_{a_0}} \frac{1}{M^2_{\rm inv}-M^2_{a_0}+i M_{a_0} \Gamma_{a_0}} \,g_{K\bar{K}} \nonumber   \\
t_{K^+ K_s^0} &=& -\frac{1}{\sqrt{2}}t_{K^+ \bar{K}^0}
\end{eqnarray}
where  in the last equation we have taken into account that $K_s^0=\frac{1}{\sqrt{2}}(K^0-\bar{K}^0)$.

We can see how ${\tilde{t}_{f_0}}$, ${\tilde{t}_{a_0}}$ appear with opposite relative signs in $K^+ K^-$ or $K^0 \bar{K}^0$ production, which explains
why there can be differences in these production rates.

Finally we have to calculate the differential decay width given by
\begin{eqnarray}\label{eq:dg}
\frac{d\Gamma}{dM_{\rm inv}(K\bar{K})}=\frac{1}{(2\pi)^3}\,\frac{1}{4 M^2_{D_s}} \, p_{\pi} \, \tilde{p}_k \, |\tilde{t_i}|^2
\end{eqnarray}
with
\begin{eqnarray}
 p_{\pi}=\frac{\lambda^{1/2}(M^2_{D_s},m^2_{\pi},M^2_{\rm inv}(K\bar{K}))}{2M_{D_s}}\,,\qquad \tilde{p}_k=\frac{\lambda^{1/2}(M^2_{\rm inv}(K\bar{K}),m^2_{K},m^2_{K})}{2M_{\rm inv}(K\bar{K})}
\end{eqnarray}
We integrate from $M_{\rm inv}=1600$ MeV to $1870$ MeV to obtain the integrated width into $\pi f_0$,$\pi a_0$  and use the data of the PDG for the  $f_0(1710)$ and from  \cite{gengvec}
for the $a_0$,
\begin{eqnarray}\label{eq:geng}
M_{f_0}&=& 1732~ {\rm MeV}\,;\quad \Gamma_{f_0}=147~ {\rm MeV} \nonumber   \\
M_{a_0}&=& 1777~ {\rm MeV}\,;\quad \Gamma_{a_0}=148~ {\rm MeV}
\end{eqnarray}

\section{Results}

As we have pointed out, the $A$, $A \alpha$, $A \beta$ parameters are associated to external emission and hence should all have a similar strength. Since $A$ disappear in ratios,
we shall take $A=1$ and then $|\alpha| \simeq |\beta| \simeq 1$. On the other hand $\gamma$ and $\delta$  come from internal emission and should have a strength of around $\frac{1}{3}$, hence
$|\gamma| \simeq |\delta| \simeq \frac{1}{3}$. But we do not know the signs. Hence we make a table of the results that we obtained using all possible signs. This makes $16$ combinations.
We define the ratio
\begin{eqnarray}
R_1=\frac{\Gamma(D_s^+ \to \pi^+ K^0 \bar{K}^0)}{\Gamma(D_s^+ \to  \pi^+ K^+ K^-)}
\end{eqnarray}
 using the results of Eqs.~\eqref{eq:ti} and \eqref{eq:dg}. Similarly we also define
\begin{eqnarray}
R_2=\frac{\Gamma(D_s^+ \to \pi^0 K^+  K_s^0)}{\Gamma(D_s^+ \to  \pi^+ K^+ K^-)}
\end{eqnarray}
and the results of $R_1$, $R_2$ are given in Table \ref{table:ratio} for the $16$ different combinations of the parameters (we integrate  $d\Gamma/dM_{\rm inv}$ over
the range of $M_{\rm inv}\in [1600-1870]~{\rm MeV}$).

\begin{center}
\begin{table*}[h]
 \renewcommand{\arraystretch}{1.}
 \setlength{\tabcolsep}{0.1cm}
 \centering
 \caption{The ratios for $16$ different sets of parameters \label{table:ratio} }
     \begin{tabular}{cccccc||cccccc}
     \hline\hline
~~$\alpha$~~~&$~~~\beta$ ~~~&~~~$\gamma$ ~~~ &~~~$\delta$ ~~~~& ~~~~  $R_1$ ~~~ &~~~ $R_2$~~~~~&~~~~$\alpha$~~&$~~~~\beta$ ~~~&~~~$\gamma$ ~~~ &~~~$\delta$ ~~~~& ~~~~  $R_1$ ~~~ &~~~ $R_2$~~~\\\hline
$1$&$1$&$\frac{1}{3}$&$\frac{1}{3}$ & $1.85$  & $ 0.15 $ & $-1$ &$1$  &$\frac{1}{3}$&$\frac{1}{3}$ & $1.53$  & $0.05 $ \\\hline
$1$ &$1$ &$\frac{1}{3}$  &$-\frac{1}{3}$ & $4.94$  & $1.44$  &$-1$&$1$&$\frac{1}{3}$&$-\frac{1}{3}$ & $4.11$  & $ 0.62 $\\\hline
$1$ &$1$&$-\frac{1}{3}$&$\frac{1}{3}$ & $2.49$  & $0.73$&$-1$&$1$&$-\frac{1}{3}$&$\frac{1}{3}$ & $1.90$  & $ 0.12$\\\hline
$1$&$1$&$-\frac{1}{3}$&$-\frac{1}{3}$ & $2.53$  & $1.51 $&$-1$&$1$&$-\frac{1}{3}$&$-\frac{1}{3}$ & $5.92$  & $1.49$\\\hline
$1$&$-1$&$\frac{1}{3}$&$\frac{1}{3}$ & $0.20 $  & $0.29$&$-1$&$-1$ &$\frac{1}{3}$&$\frac{1}{3}$ & $0.24$  & $0.15 $\\\hline
$1$&$-1$&$\frac{1}{3}$&$-\frac{1}{3}$ & $0.54 $  & $0.08$&$-1$&$-1$&$\frac{1}{3}$&$-\frac{1}{3}$ & $0.66$  & $0.03 $\\\hline
$1$&$-1$&$-\frac{1}{3}$&$\frac{1}{3}$ & $0.40 $  & $0.60$&$-1$&$-1$&$-\frac{1}{3}$&$\frac{1}{3}$ & $0.17 $  & $0.25 $\\\hline
$1$&$-1$&$-\frac{1}{3}$&$-\frac{1}{3}$ & $0.40 $  & $0.29 $ &$-1$&$-1$&$-\frac{1}{3}$&$-\frac{1}{3}$ & $0.53 $  & $0.06$\\
\hline \hline
\end{tabular} 
\end{table*}
\end{center}

We should compare these results with experiment. Recalling the discussion in Ref. \cite{besnew},
we see that ${\mathrm{Br}}[D_s^+ \to  \pi^+ f_0(1710)]$ from \cite{besold} is \footnote{Recall that the message in Ref. \cite{besnew} is that both in \cite{besnew} and \cite{besold}
what is assumed  $f_0(1710)$  is actually a combination of  $f_0(1710)$  and  $a_0(1710)$. This is what mean by `` ". }
\begin{equation}\label{eq:oldBr}
 {\mathrm{Br}}[D_s^+ \to  \pi^+ ``f_0(1710)";~``f_0(1710)" \to K^+ K^-]=(1.0 \pm 0.2\pm 0.3)\times 10^{-3}
\end{equation}
and from \cite{besnew}
\begin{equation}\label{eq:newBr}
 {\mathrm{Br}}[D_s^+ \to  \pi^+ ``f_0(1710)";~ ``f_0(1710)" \to K_s^0 K_s^0]=(3.1 \pm 0.3\pm 0.1)\times 10^{-3}
\end{equation}
This gives
\begin{eqnarray}\label{eq:expR1}
R_1=2\times\frac{\Gamma(D_s^+ \to \pi^+ K_s^0 K_s^0)}{\Gamma(D_s^+ \to  \pi^+ K^+ K^-)} =6.20 \pm 0.67
\end{eqnarray}
where we have added experimental errors in quadrature. We have a bracket of experimental values of $R_1 \in [5.57-6.83]$.

  From Table \ref{table:ratio} we observe that among all the $16$ possible combinations of parameters only two
  give results for $R_1$ comparable to those of Eq.~\eqref{eq:expR1}.   These are
\begin{align} \label{eq:start}
\begin{cases}
 \alpha=1,\beta=1,\gamma=\frac{1}{3},\delta=-\frac{1}{3}  & \bl{(\rm set ~I)} \\[0.1cm]
 \alpha=-1,\beta=1,\gamma=-\frac{1}{3},\delta=-\frac{1}{3}  & \bl{(\rm set ~II)}
 \end{cases}
\end{align}
with some preference for the second set which gives $R_1$ compatible with experiment. In both cases we see that the ratio of $R_2$ is of the order
of $1.4$, indicating a branching ratio for
\begin{equation}
 {\mathrm{Br}}[D_s^+ \to  \pi^0 a_0(1710); a_0(1710)\to K^+ K_s^0] \simeq 1.4\times 10^{-3}
\end{equation}
There are yet no data for this decay which is in the process of analysis, as indicated in \cite{besnew}. This number
should be considered a neat prediction of our approach, together with the fact that the large ratio of $R_1$ observed in  \cite{besnew}
finds a natural explanation within our theoretical  framework.

  So far we have taken sharp values for  $\alpha,\beta,\gamma,\delta$ and the results obtained are reasonably close to experimental values.
We can make small changes in the parameters  to reach values closer to the  experiment. For this we take the solutions of set I and set II and
make small increments of the parameters in the direction of the gradient of $R_1$,
$\overrightarrow{\nabla} R_1\equiv (\frac{\partial R_1}{\partial \alpha_i};\alpha_1=\alpha, \alpha_2=\beta,\alpha_3=\gamma,\alpha_4=\delta)$ till we find
a value in the range of the experimental value of Eq.~\eqref{eq:expR1}. With these parameters we evaluate $R_2$. We find the following results
\begin{align}\label{eq:draw}
\begin{cases}
 R_1=5.95\,,\quad R_2=1.88 ~& \bl{\rm set ~I:~} \alpha=0.739,\beta=0.764,\gamma=0.483,\delta=-0.783\\[0.1cm]
 R_1=6.23\,,\quad R_2=2.02 ~& \bl{\rm set ~II:~} \alpha=-0.996,\beta=1.011,\gamma=-0.361,\delta=-0.382
 \end{cases}
\end{align}
The values of the parameters have changed little and then we still can conclude that with natural values of the parameters for the external and
internal emission we obtain results for the $R_1$ ratio compatible with the experiment of \cite{besnew}. We can not decide on which of the two solutions
is the correct one, although the proximity of the parameters of set II to the starting ones of Eq.~\eqref{eq:start}  reinforces our preference for this set.
 Interestingly, the predictions for the $R_2$  ratio are very similar with two sets. We can safely say that our approach predicts a ratio
\begin{eqnarray}
R_2 \in [1.8-2.0]
\end{eqnarray}
or what is the same, using the experimental data of  Eq.~\eqref{eq:oldBr}, and summing errors in quadrature,
\begin{equation}\label{eq:nR2}
 {\mathrm{Br}}[D_s^+ \to  \pi^0 a_0(1710); a_0(1710)\to K^+ K_s^0] =(2.0 \pm 0.7) \times 10^{-3}
\end{equation}
This is a neat prediction of our approach and it will be most interesting to see what the coming experiment mentioned in \cite{besnew} reports.

\begin{figure}[h]
\centering
\includegraphics[scale=0.9]{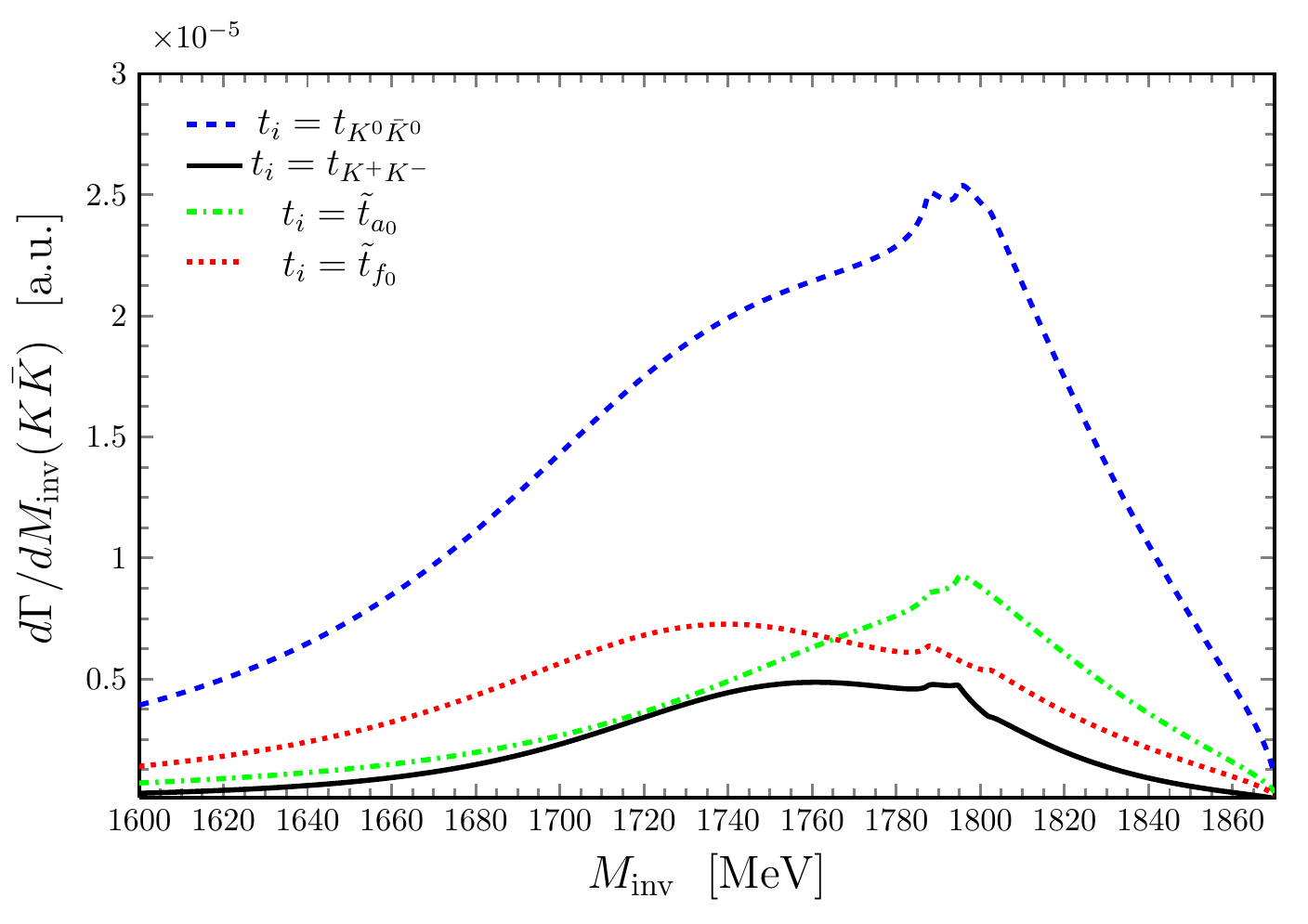}
\caption{Mass distributions $d\Gamma/dM_{\rm inv}$ for the cases of Eq.~\eqref{eq:tii}. The result for $K^+ K_s^0$ is the same as for the
case of ${\tilde{t}_{a_0}}$ in the figure. The results correspond to the value of the parameters of set II of Eq.~\eqref{eq:draw} .}
\label{fig:pc}
\end{figure}

To finalize our study we plot in Fig.~\ref{fig:pc} the results of $d\Gamma/dM_{\rm inv}(K\bar{K})$  of Eq.~\eqref{eq:dg} with the
different amplitudes $t_i$ of Eq.~\eqref{eq:ti}, with
\begin{align}\label{eq:tii}
\begin{cases}
{\rm  1)}~  t_i={\tilde{t}_{f_0}} {\rm ~part~ of~} t_{K^+ K^-} \\[0.1cm]
{\rm  2)}~  t_i={\tilde{t}_{a_0}} {\rm ~part~ of~} t_{K^0 \bar{K}^0} \\[0.1cm]
{\rm  3)}~  t_i=t_{K^+ K^-}  \\[0.1cm]
{\rm  4)}~ t_i=t_{K^0 \bar{K}^0}
 \end{cases}
\end{align}

As we can see, the differential mass distribution for the case of only the  $a_0(1710)$ contribution has a strength a bit bigger than the one
with only the  $f_0(1710)$ contribution. The peak of the $a_0$ contribution is displaced to the right relative to the $f_0$ contribution
because we use the theoretical mass of \cite{gengvec} [see Eq.~\eqref{eq:geng}]. This distribution should be the same seen in the
$D_s^+ \to \pi^0 K^+ K_s^0$ reaction. It will be most interesting to see the peak position in the coming experiment of the
$D_s^+ \to \pi^0 K^+ K_s^0$ decay. Note that uncertainties of $30-40$ MeV in the predicted position of the resonance are normal in the approach of \cite{gengvec}.

What is  clear from the figure is that in the $K^0 \bar{K}^0$ mass distribution there has been a constructive interference of the $f_0$ and
$a_0$ resonances, while in the $K^+ K^-$ mass distribution the interference has been destructive. This is exactly the reason given in
the analysis of the $D_s^+ \to  \pi^+ K^+ K^-$ and $D_s^+ \to  \pi^+ K_s^0 K_s^0 $ reactions \cite{besnew}
to justify the existence of the $a_0(1710)$ resonance, which should give the same $K^+ K^-$ or  $K^0 \bar{K}^0$ mass distributions should there
be only the $f_0(1710)$ state.

\section{Conclusions}
The appearance of the  $D_s^+ \to \pi^+ K_s^0 K_s^0$ experiment \cite{besnew} contrasting the results with those observed in the
$D_s^+ \to  \pi^+ K^+ K^-$ reaction in \cite{besold} and claiming the existence of an $a_0$ resonance around $1710$ MeV, the isospin
parter of the $f_0(1710)$ (with mass $1735$ MeV in the PDG), motivated us to perform this  work, since indeed such a resonance had been
predicted in Ref. \cite{gengvec}  as  a molecular state of $K^* \bar{K}^*$ and other vector-vector coupled channels.

 We looked into the possible ways that the  $f_0(1710)$  and  $a_0(1710)$ could be produced in the
 $D_s^+ \to  \pi^+ K^+ K^-,  \pi^+  K^0 \bar{K}^0,  \pi^0  K^+ \bar{K}^0$ decays and we identified five different modes in which they could be produced.
 Three of them associated with external emission, to which we gave weights $A, A\alpha, A \beta$ and two modes associated with internal emission
 with weights $A\gamma, A \delta$. The value of $A$ is irrelevant since it is related to the global strength and disappears when
 we perform ratios of rates. While it might look like we have four parameters free, this is not the case, since taking $A=1$, $\alpha$ and $\beta$,
 corresponding to external emission like the case $A$, will also have weight around $1$, but with unknown sign. Similarly, the $\gamma$ and $\delta$
 parameters  corresponding to decay modes of internal emission, which are suppressed by a color factor $N_c$, will have a weight around $\frac{1}{3}$,
 again with  unknown signs. We calculated the ratio of the  $D_s^+ \to \pi^+  K^0 \bar{K}^0$ and $D_s^+ \to  \pi^+ K^+ K^-$ decay widths with the $16$
possible sign combinations with the strength of the parameters discussed above, and found that only two of them gave an acceptable  value of the ratios
compared with experiment. A fine tuning  of the parameters with small deviations from the standard values gave us values of the ratio in agreement with the
range of the experimental value found in  \cite{besnew}.  In this way we have shown that the picture of  \cite{gengvec}  for the $a_0(1710)$ state
provides results for this ratio in  agreement with the findings of the experiment.

 Another interesting result of our study is that we make predictions for the branching ratio of the, yet, unknown results for the
 $D_s^+ \to \pi^0 K^+ K_s^0$ reaction. For either of the two sets of parameters that we found acceptable we found branching ratios for this reaction
 in the range of $(2.0 \pm 0.7)\times 10^{-3}$. This is a neat prediction of our theoretical approach which is only tied to the theoretical couplings
 of the $f_0(1710)$  and  $a_0(1710)$ resonances found in  \cite{gengvec} to the different coupled channels that build up the resonance, and to the
 experimental value of the ratio of branching ratios of  $D_s^+ \to \pi^+ K_s^0 K_s^0$ and $D_s^+ \to \pi^0 K^+ K^-$ found in \cite{besnew}.
 An agreement of the coming results of the $D_s^+ \to \pi^0 K^+ K_s^0$  reaction with the predictions made here would give a boost to the molecular interpretation
 on the nature of these two resonances, while a deviation by one order of magnitude would certainly pose serious trouble to that picture.
 The present work makes then very valuable the expected results for the   $D_s^+ \to \pi^0 K^+ K_s^0$  reaction.

\section*{ACKNOWLEDGEMENT}
This work is partly supported by the National Natural Science Foundation of China under Grants Nos. 11975009, 12175066 and No. 11975041.
This work is also partly supported by the Spanish Ministerio de Economia y Competitividad (MINECO) and European FEDER funds
under Contracts No. FIS2017-84038-C2-1-P B, PID2020-112777GB-I00, and by Generalitat Valenciana under contract PROMETEO/2020/023.
This project has received funding from the European Union Horizon 2020 research and innovation programme under
the program H2020-INFRAIA-2018-1, grant agreement No. 824093 of the ¡°STRONG-2020¡± project.



\begin{thebibliography}{}

\bibitem{pdg}
P.A. Zyla et al. (Particle Data Group), Prog. Theor. Exp. Phys. \textbf{2020}, 083C01 (2020)

\bibitem{isgur}
S. Godfrey and N. Isgur, Phys. Rev. D \textbf{32}, 189 (1985)

\bibitem{entem} J. Segovia, D. R. Entem, F. Fern\'{a}ndez, Phys. Lett. B \textbf{662}, 33 (2008)

\bibitem{vijande} J.~Vijande, F.~Fernandez and A.~Valcarce, J. Phys. G \textbf{31}, 481 (2005)


\bibitem{gengvec} L. S. Geng and E. Oset, Phys. Rev. D \textbf{79}, 074009 (2009)

\bibitem{raquelvec} R. Molina, D. Nicmorus, and E. Oset, Phys. Rev. D \textbf{78}, 114018 (2008)

\bibitem{hidden1} M. Bando, T. Kugo, K. Yamawaki,  Phys. Rept. \textbf{164}, 217 (1988)

\bibitem{hidden2} M. Harada, K. Yamawaki,  Phys. Rept. \textbf{381}, 1 (2003)

\bibitem{hidden4} Ulf-G. Meissner,  Phys. Rept. \textbf{161}, 213 (1988)

\bibitem{hideko} H. Nagahiro, L. Roca, A. Hosaka, and E. Oset, Phys. Rev. D \textbf{79}, 01401  (2009)

\bibitem{genghideko} T. Branz, L. S. Geng, and E. Oset, Phys. Rev. D \textbf{81}, 054037 (2010)

\bibitem{albeto} A. Mart\'{i}nez Torres, K. P. Khemchandani, F. S. Navarra, M. Nielsen, E. Oset,  Phys. Lett. B \textbf{719}, 388 (2013)

\bibitem{expfiom} M. Ablikim et al. (BES Collaboration),  Phys. Rev. Lett. \textbf{96}, 162002 (2006)

\bibitem{gengguo} L. S. Geng, F. K. Guo, B. S. Zou, Eur. Phys. J. A \textbf{44}, 305 (2010)

\bibitem{yamagata} J. Yamagata-Sekihara, E. Oset,  Phys. Lett. B \textbf{690}, 376 (2010)

\bibitem{jujun} J. J. Xie and E. Oset, Phys. Rev. D \textbf{90}, 094006 (2014)

\bibitem{daijujun} L. R. Dai, J. J. Xie, and E. Oset, Phys. Rev. D \textbf{91}, 094013 ( 2015)

\bibitem{daigeng} R. Molina, L. R. Dai,  L. S.Geng,  and E. Oset,  Eur. Phys. J. A \textbf{56}, 173 (2020)

\bibitem{ikenoliang} N. Ikeno, J. M. Dias, W. H. Liang, and E. Oset,  Phys. Rev. D \textbf{100}, 114011 (2019)

\bibitem{lees} J. P. Lees et al. (BABAR Collaboration), Phys. Rev. D \textbf{104}, 072002 (2021)


\bibitem{besnew} M. Ablikim et al. (BESIII Collaboration), 	arXiv: 2110.07650 [hep-ex]

\bibitem{besold} M. Ablikim et al. (BESIII Collaboration), Phys. Rev. D \textbf{104}, 012016 (2021)


\bibitem{chau} L. L. Chau, Phys. Rept. \textbf{95}, 1 (1983)

\bibitem{xiao} Z. Y. Wang, J. Y. Yi, Z. F. Sun, C. W. Xiao, arXiv: 2109.00153 [hep-ph]

\bibitem{bramon} A. Bramon, A. Grau, G. Pancheri, Phys. Lett. B \textbf{283}, 416 (1992)

\end{thebibliography}
\end{document}